\documentclass[twocolumn,english,aps,prb,twocolum,superscriptaddress,bibnotes,amsmath,amssymb,floatfix,footinbib]{revtex4-2}

\usepackage{amsmath,amsfonts,amssymb}
\usepackage[english]{babel}
\usepackage[markup=blue authormarkupposition=left]{changes}
\usepackage{color}
\usepackage[T1]{fontenc}
\usepackage[colorlinks=true,citecolor=blue,linkcolor=magenta]{hyperref}
\usepackage[utf8]{inputenc}
\usepackage{epstopdf}
\usepackage{graphicx}
\usepackage{mhchem}
\usepackage{physics}
\usepackage{siunitx}
\usepackage{soul}
\usepackage{upgreek}
\usepackage{url}

\graphicspath{{./Figures/}}

\newcommand{\EPFL}{Institute of Physics, Swiss Federal Institute of Technology Lausanne (EPFL), CH-1015 Lausanne, Switzerland}
\newcommand{\QSE}{Center for Quantum Science and Engineering, Swiss Federal Institute of Technology Lausanne (EPFL), CH-1015 Lausanne, Switzerland}

\begin{document}

\title{Unifying frequency metrology across microwave, optical, and free-electron domains}

\author{Yujia Yang}
\thanks{These authors contributed equally.}
\affiliation{\EPFL}
\affiliation{\QSE}
\author{Paolo Cattaneo}
\thanks{These authors contributed equally.}
\affiliation{\EPFL}
\author{Arslan S. Raja}
\affiliation{\EPFL}
\affiliation{\QSE}
\author{Bruce Weaver}
\affiliation{\EPFL}
\author{Rui Ning Wang}
\affiliation{\EPFL}
\affiliation{\QSE}
\author{Alexey Sapozhnik}
\affiliation{\EPFL}
\author{Fabrizio Carbone}
\affiliation{\EPFL}
\author{Thomas LaGrange}
\email{thomas.lagrange@epfl.ch}
\affiliation{\EPFL}
\author{Tobias J. Kippenberg}
\email{tobias.kippenberg@epfl.ch}
\affiliation{\EPFL}
\affiliation{\QSE}

\maketitle

%%%%%%%%%%%%%%%%%%%%%%%%%%%%%%%%%%%%%%%%%%%%%%%%%%%%%%%%%%%%%%%%%%%%%%
%%%%%%%%%%%%%%%%%%%%%%%%%%%%% Abstract %%%%%%%%%%%%%%%%%%%%%%%%%%%%%%%
%%%%%%%%%%%%%%%%%%%%%%%%%%%%%%%%%%%%%%%%%%%%%%%%%%%%%%%%%%%%%%%%%%%%%%
\noindent\textbf{Frequency metrology lies at the heart of precision measurement.
Optical frequency combs provide a coherent link uniting the microwave and optical domains in the electromagnetic spectrum, with profound implications in timekeeping~\cite{diddamsOpticalClockBased2001, ludlowOpticalAtomicClocks2015}, sensing and spectroscopy~\cite{holzwarthOpticalFrequencySynthesizer2000, diddamsMolecularFingerprintingResolved2007, picqueFrequencyCombSpectroscopy2019}, fundamental physics tests~\cite{bizeTestingStabilityFundamental2003, rosenbandFrequencyRatioHg2008}, exoplanet search~\cite{steinmetzLaserFrequencyCombs2008a, liLaserFrequencyComb2008b, wilkenSpectrographExoplanetObservations2012}, and light detection and ranging~\cite{minoshimaHighaccuracyMeasurement240m2000, coddingtonRapidPreciseAbsolute2009, trochaUltrafastOpticalRanging2018a}. 
Here, we extend this frequency link to free electrons by coherent modulation of the electron phase by a continuous-wave laser locked to a fully stabilized optical frequency comb.
Microwave frequency standards are transferred to the optical domain via the frequency comb, and are further imprinted in the electron spectrum by optically modulating the electron phase with a photonic chip-based microresonator.
As a proof-of-concept demonstration, we apply this frequency link in the calibration of an electron spectrometer, and use the electron spectrum to measure the optical frequency. 
Our work bridges frequency domains differed by a factor of $\sim10^{13}$ and carried by different physical objects, establishes a spectroscopic connection between electromagnetic waves and free-electron matter waves, and has direct ramifications in ultrahigh-precision electron spectroscopy.
}

%%%%%%%%%%%%%%%%%%%%%%%%%%%%%%%%%%%%%%%%%%%%%%%%%%%%%%%%%%%%%%%%%%%%%%
%%%%%%%%%%%%%%%%%%%%%%%%%%%%%%%%%%%%%%%%%%%%%%%%%%%%%%%%%%%%%%%%%%%%%%
%%%%%%%%%%%%%%%%%%%%%%%%%%%%%%%%%%%%%%%%%%%%%%%%%%%%%%%%%%%%%%%%%%%%%%

%%%%%%%%%%%%%%%%%% Fig. 1 %%%%%%%%%%%%%%%%%%
\begin{figure*}[ht]
\centering
\includegraphics[width=\textwidth]{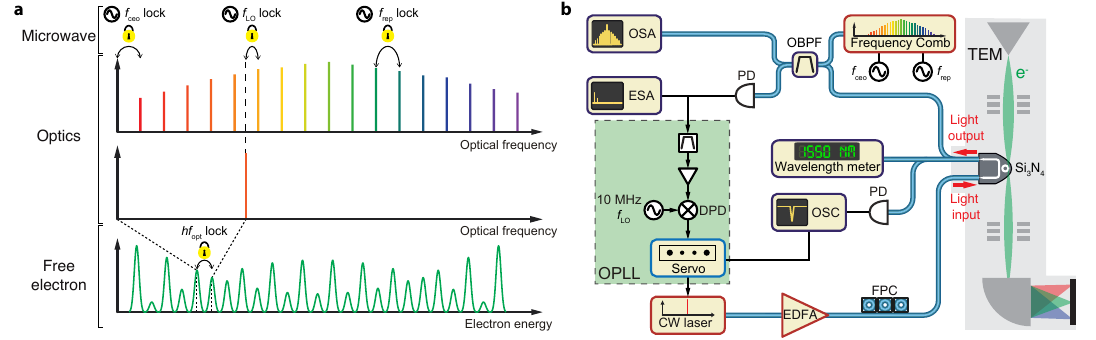}
\caption{
\textbf{Uniting optical frequency metrology with electron spectroscopy.}
\textbf{a}, Concept of frequency metrology across microwave, optical, and free-electron domains. Microwave and optical domains are connected by a fully stabilized optical frequency comb with the carrier-envelope-offset frequency $f_{\mathrm{ceo}}$ and the repetition rate $f_{\mathrm{rep}}$ being locked. A continuous-wave (CW) laser is offset locked to one comb tooth via a local oscillator (LO) at frequency $f_{\mathrm{LO}}$. The CW laser then coherently modulates the phase of an electron beam, generating sidebands in the electron spectrum that are regularly spaced by the photon energy $hf_{\mathrm{opt}}$.  
\textbf{b}, Experimental setup. A CW laser pumps a \ce{Si3N4} photonic chip-based microresonator in a transmission electron microscope (TEM) to modulate the electron phase and broaden the electron spectrum measured in a post-column electron spectrometer. A wavelength meter performs coarse measurement of the laser frequency. The transmitted CW laser is mixed with a stabilized optical frequency comb (OFC) to generate microwave beatnotes that are measured with an electronic spectrum analyzer (ESA). An optical spectrum analyzer (OSA) monitors the laser and the OFC. An optical phase-locked loop (OPLL) offset-locks the laser to one comb tooth. The beatnote and a \SI{10}{\MHz} microwave local oscillator (LO) are mixed at a 12-bit double-balanced digital phase detector (DPD) that accounts for large phase slips induced by frequency fluctuations. The DPD output, or error signal, is fed to a servo controller to adjust the laser frequency. Blue paths represent fiber connections, and black lines stand for electronic connections. EDFA: erbium-doped fiber amplifier; FPC: fiber polarization controller; PD: photodetector; OSC: oscilloscope; OBPF: optical bandpass filter.
}
\label{Fig:setup}
\end{figure*}
%%%%%%%%%%%%%%%%%% Fig. 1 %%%%%%%%%%%%%%%%%%

Frequency is the most accurately measured physical quantity and is pivotal to precision metrology in a multitude of scientific and technological sectors.
It was advised by Arthur Schawlow, the Nobel laureate in physics in 1981, to ``never measure anything but frequency''~\cite{hanschNobelLecturePassion2006a}.
The optical frequency comb (OFC) consists of a `comb' of precisely equidistant frequency lines at optical frequencies separated by a microwave frequency~\cite{udemOpticalFrequencyMetrology2002, diddamsOpticalFrequencyCombs2020}.
The OFC serves as a phase-coherent link, or a `light gear', that accurately connects frequency components in the microwave and optical realms, and has revolutionized a plethora of fields including atomic clocks~\cite{diddamsOpticalClockBased2001, ludlowOpticalAtomicClocks2015}, spectroscopy~\cite{holzwarthOpticalFrequencySynthesizer2000, diddamsMolecularFingerprintingResolved2007, picqueFrequencyCombSpectroscopy2019}, ultrafast optics~\cite{cundiffColloquiumFemtosecondOptical2003}, telecommunications~\cite{marin-palomoMicroresonatorbasedSolitonsMassively2017b}, navigation and ranging~\cite{minoshimaHighaccuracyMeasurement240m2000, coddingtonRapidPreciseAbsolute2009, trochaUltrafastOpticalRanging2018a}.

Electron spectroscopy uses free electrons to perform spectroscopic measurements of specimens.
In particular, electron energy-loss spectroscopy (EELS) measures the energy loss of electrons transmitting through and inelastically scattered by a specimen, based on the spatial distribution of the electron beam bent and dispersed by a magnetic field~\cite{egertonElectronEnergyLossSpectroscopy2011, egertonElectronEnergylossSpectroscopy2008a}.
Typically implemented in a transmission electron microscope (TEM) with a post-column spectrometer, EELS possesses sub-atomic spatial resolution and is a powerful probe for local excitations or chemically resolved imaging, widely applied in the investigation of atomic composition~\cite{mullerAtomicScaleChemicalImaging2008}, chemical bonding~\cite{batsonSimultaneousSTEMImaging1993, tanOxidationStateChemical2012}, and vibrational properties~\cite{krivanekVibrationalSpectroscopyElectron2014, lagosMappingVibrationalSurface2017} of materials, macromolecule assemblies and subcellular compartments in biological systems~\cite{aronovaDevelopmentElectronEnergyloss2012}, as well as thermal~\cite{mecklenburgNanoscaleTemperatureMapping2015}, electronic~\cite{martinAnalyzingChannelDopant2018}, and optical~\cite{nelayahMappingSurfacePlasmons2007a} properties of nanoscale devices.
Despite the excellent spatial resolution, the spectral resolution of EELS (about \SI{e-3}{\eV} for state-of-the-art techniques~\cite{krivanekProgressUltrahighEnergy2019b}) is far inferior to that of its optical counterparts, where the optical frequency comb allows measuring absolute frequencies with precision reaching mHz-level or \SI{e-18}{\eV}~\cite{hutsonObservationMillihertzlevelCooperative2024}.
Moreover, the frequency reference or calibration standard for EELS relies on elemental edges associated with inner shell ionizations.
This type of calibration has a limited accuracy and applicable spectral range, and needs to presume a linear dispersion. 
Here, we unite electron spectroscopy with optical frequency metrology, and use the precisely measured photon energy for EELS calibration.

\section{Frequency metrology across microwave, optical, and free-electron domains}
Figure \ref{Fig:setup}a illustrates the concept of frequency metrology uniting microwave, optical, and free-electron domains. 
The microwave and optical frequencies are linked by an optical frequency comb, whose carrier-envelope-offset frequency $f_{\mathrm{ceo}}$ and repetition rate $f_{\mathrm{rep}}$ are locked.
Therefore, the comb teeth are located at optical frequencies that are precisely synthesized from microwave frequencies: $	f_{\mathrm{m}} = f_{\mathrm{ceo}} + m \times f_{\mathrm{rep}}$, with $m$ the mode number.
A monochromatic continuous-wave (CW) laser is then offset-locked to one comb tooth with a microwave local oscillator (LO), and one obtains a synthesized optical frequency
%mixed with the frequency comb at a fast photodetector, generating beatnotes at microwave frequencies that can be precisely measured for determining the absolute optical frequency of the CW laser.
%With a feedback loop, the laser frequency can be stabilized by locking the frequency of one beatnote to a microwave local oscillator (LO).
%As a result, one could obtain a synthesized optical frequency
\begin{equation}\label{eq:f_opt}
	f_{\mathrm{opt}} = f_{\mathrm{ceo}} + m \times f_{\mathrm{rep}} \pm f_{\mathrm{LO}}.
\end{equation}
%Note that the ambiguity ($\pm$) can be removed by observing the shift direction of the beatnote while adjusting the laser frequency. 

The CW laser is then used to impose a coherent phase modulation onto free-electron wavefunctions~\cite{barwickPhotoninducedNearfieldElectron2009b, garciadeabajoMultiphotonAbsorptionEmission2010, parkPhotoninducedNearfieldElectron2010}, which broadens the initially narrow electron spectrum into a comb-like structure consisting of regularly spaced energy sidebands (see SI) at
\begin{equation}\label{eq:E_N}
	E_{\mathrm{N}} = E_0 + N \times hf_{\mathrm{opt}},
\end{equation}
where $E_0$ is the initial electron energy, $h$ is the Planck constant, and $N$ is an integer.
Equally, the generation of energy sidebands can be understood as quantum-mechanical superpositions of electron states corresponding to the absorption or emission of $N$ photons in the inelastic electron-light scattering (IELS) process.
In the end, one obtains an electron spectrum consisting of energy sidebands centered at the initial electron energy $E_0$ with energy change
\begin{equation}\label{eq:E_N_E_0}
	E_{\mathrm{N}} - E_0 = N h (f_{\mathrm{ceo}} + mf_{\mathrm{rep}} \pm f_{\mathrm{LO}}),
\end{equation}
arising from electron phase modulation at an optical frequency synthesized from microwave frequencies.

%%%%%%%%%%%%%%%%%% Fig. 2 %%%%%%%%%%%%%%%%%%
\begin{figure*}[ht]
\centering
\includegraphics[width=\textwidth]{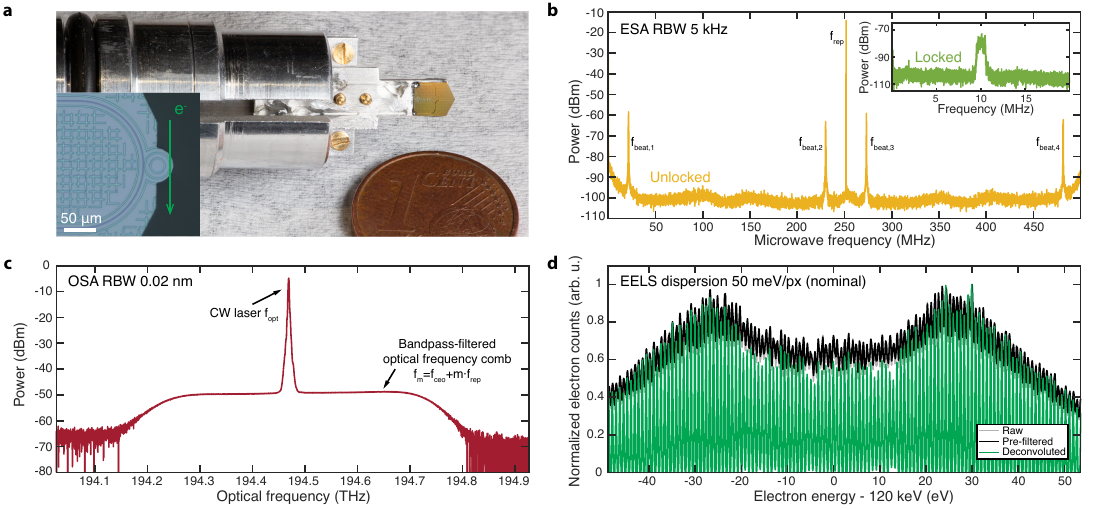}
\caption{
\textbf{Frequency-locked microwave, optical, and electron spectra}. 
\textbf{a}, Photograph of the photonic chip mounted on a custom holder. Inset: optical micrograph of the microresonator and schematic depiction of the electron beam position.
\textbf{b}, Microwave spectrum recorded with an ESA. The beatnote around \SI{250}{\MHz} is from the repetition rate $f_{\mathrm{rep}}$ of the optical frequency comb, and the other beatnotes are generated by beating the CW laser with the frequency comb teeth. Inset: microwave spectrum when the OPLL is activated.
\textbf{c}, Optical spectrum recorded with an OSA. The narrow peak corresponds to the CW laser and the lower plateau is from the bandpass-filtered frequency comb. Individual comb lines are not resolved by the OSA. 
\textbf{d}, Electron spectrum recorded with the electron spectrometer. The raw spectrum (gray) is first smoothed (black) with a Gaussian pre-filter, and then deconvoluted to generate a comb-like spectrum with well-separated peaks (green).  
}
\label{Fig:spectra}
\end{figure*}
%%%%%%%%%%%%%%%%%% Fig. 2 %%%%%%%%%%%%%%%%%%

We use a high-quality-factor \ce{Si3N4} photonic chip-based microresonator~\cite{pfeifferPhotonicDamasceneProcess2016a, liuHighyieldWaferscaleFabrication2021} driven by a CW laser to implement the electron phase modulation (Fig. \ref{Fig:setup}b).
Though coherent optical modulation of free electrons was achieved more than a decade ago in the context of photon-induced near-field electron microscopy~\cite{barwickPhotoninducedNearfieldElectron2009b}, it is the development of photonic integrated circuit-based electron phase modulation~\cite{henkeIntegratedPhotonicsEnables2021c} that permits both CW laser-driven operation and strong phase modulation.
These attributes translate into a narrow-linewidth driving laser with a well-defined frequency $f_{\mathrm{opt}}$ and a large electron spectral broadening, both important to the application of precise optical frequency metrology in electron spectroscopy.

%%%%%%%%%%%%%%%%%% Fig. 3 %%%%%%%%%%%%%%%%%%
\begin{figure*}[ht]
\centering
\includegraphics[width=\textwidth]{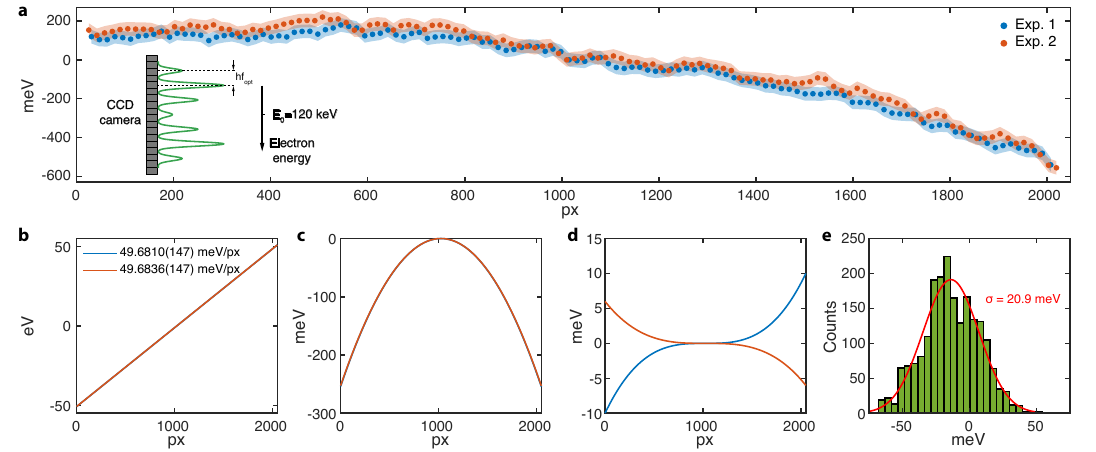}
\caption{
\textbf{Electron spectrometer calibration with two independent exposures}.
\textbf{a}, Deviations of the calibration markers at multiples of the photon energy from the nominal energy of the spectrometer for Exposures (Exp.) 1 and 2. Shaded areas correspond to error bars. Inset: a schematic depiction of obtaining the energy solution from the electron spectrum.
\textbf{b}, \textbf{c}, \textbf{d}, First, second, and third-order components of the polynomial fits of the energy solutions (pixel-to-energy mappings) for the two exposures. The first-order components illustrate the calibrated average dispersion.
\textbf{e}, Histogram of the pixel-wise root-mean-square deviation between the energy solutions of the two exposures. Red curve represents a normal distribution fit with a standard deviation of \SI{20.9}{\meV}.
}
\label{Fig:2exp}
\end{figure*}
%%%%%%%%%%%%%%%%%% Fig. 3 %%%%%%%%%%%%%%%%%%

The experimental setup (Fig. \ref{Fig:setup}b) consists of a TEM with a post-column electron energy-loss spectrometer, a photonic chip-based microresonator mounted on a custom holder (Fig. \ref{Fig:spectra}a), and an optical frequency synthesizer based on a frequency comb. 
Efficient interaction requires the electron velocity to match the phase velocity of light (phase-matching condition). 
In the present case, electrons ($\sim\SI{120}{\keV}$ energy, corresponding to $\sim\SI{2.9e19}{\Hz}$ frequency) phase-matched to the optical wave ($\sim\SI{1550}{\nm}$ wavelength) traverse the evanescent near-field of the microresonator and are spectrally broadened by the IELS~\citep{henkeIntegratedPhotonicsEnables2021c}.
An optical phase-locked loop (OPLL) locks the CW laser to one tooth of a stabilized OFC with a \SI{10}{\MHz} offset frequency defined by a microwave LO, so that the absolute laser frequency is precisely determined.
Figure \ref{Fig:spectra}b shows a typical microwave spectrum with multiple beatnotes.
The beatnote near \SI{250}{\MHz} originates from the repetition rate $f_{\mathrm{rep}}$ of the OFC, while the other 4 beatnotes are generated by beating the CW laser with the comb teeth of the OFC.
The absolute optical frequency can be precisely measured from the microwave beatnote and the mode number $m$ determined by the wavelength meter.
When the OPLL is activated, the beatnote with the lowest frequency is locked to the LO at $f_{\mathrm{LO}}=\SI{10}{\MHz}$, and the microwave spectrum is displayed in the inset of Fig. \ref{Fig:spectra}b.
In this way, an absolute optical frequency is synthesized.
Figure \ref{Fig:spectra}c illustrates the corresponding optical spectrum.
Note that OSA only monitors the optical frequency components, while precise frequency measurement is performed in the microwave domain.
The optical spectrum features a narrow peak from the CW laser, and a plateau from the bandpass-filtered OFC (individual comb lines cannot be resolved).

Figure \ref{Fig:spectra}d demonstrates an example of the measured electron spectrum generated by the IELS.
Due to the finite spectral width ($\SI{0.7}-\SI{0.8}{\eV}$) of the incident electrons and hence the elastically scattered electrons (known as the zero-loss peak, ZLP), the energy sidebands at multiples of the photon energy ($\sim\SI{0.8}{\eV}$) are not fully separated and overlap each other.
To mitigate this issue, we first smooth the raw data by convolution with a Gaussian pre-filter, and then use the Richardson-Lucy deconvolution~\cite{gloterImprovingEnergyResolution2003} to retrieve a comb-like spectrum with well-separated peaks.
These electron spectral peaks are spaced by a photon energy $hf_{\mathrm{opt}}$, with the absolute optical frequency $f_{\mathrm{opt}}$ precisely measured or synthesized from microwave frequencies $f_{\mathrm{rep}}$, $f_{\mathrm{ceo}}$, and $f_{\mathrm{LO}}$ (cf. Eq. \ref{eq:E_N_E_0}).
Consequently, a frequency link is built across microwave, optical, and free-electron domains.

\section{Ultrahigh-precision calibration of an electron spectrometer}
As a proof-of-concept demonstration, we apply the aforementioned frequency metrology to the calibration of an EELS spectrometer attached to an un-monochromated TEM with a thermionic electron gun.
Recent advances in electron monochromators, spectrometers, and detectors have pushed the energy resolution of EELS to the millielectronvolt level~\cite{krivanekProgressUltrahighEnergy2019b}.
However, the typical calibration methods of electron spectrometers fall short of the required precision or demand complicated operations.
The standard calibration based on ionization edges~\cite{potapovMeasuringAbsolutePosition2004a} has a limited spectral range and accuracy and is insensitive to local changes of dispersion.
The drift tube scanning method~\cite{websterCorrectionEELSDispersion2020a} could perform local calibration, but suffers from multiple acquisitions and voltage instabilities.

The frequency metrology across microwave, optical, and free-electron domains provides a direct frequency link from precisely measured microwave frequencies to the change of the electron energy (cf. Eq. \ref{eq:E_N_E_0}).
Therefore, the IELS-broadened electron spectrum serves as an energy reference for EELS calibration, using the energy sidebands located at multiples of the photon energy $hf_{\mathrm{opt}}$ as calibration markers.

We employ the IELS-broadened electron spectrum in calibrating an electron spectrometer with 2048 pixels set at \SI{50}{\meV/px} nominal dispersion (Fig. \ref{Fig:2exp}).
The goal of the calibration is to obtain an `energy solution', which is a precise pixel-to-energy mapping that might differ from the nominal dispersion specified by the manufacturer.
Pixel positions (with sub-pixel precision) for the calibration markers are assigned an energy of the corresponding multiples of the photon energy; pixels in between the markers are assigned an energy from linear interpolation.
We perform the calibration with two independent exposures, both using electrons modulated by a CW laser at a frequency of \SI{194 567 516}{\MHz}.
Figure \ref{Fig:2exp}a displays the deviation of the calibration markers (at multiples of the photon energy) from the nominal energy that is linear with the pixel number with a slope of \SI{50}{\meV/px}.
Both calibrations show a negative deviation indicating the nominal dispersion overestimates the true dispersion and that the common assumption of having a constant dispersion is inaccurate.
Figures \ref{Fig:2exp}b-d depict the first, second, and third-order components of the polynomial fits of the energy solutions for the two exposures.
The first-order components represent the calibrated average dispersion, which is $\sim\SI{300}{\micro\eV/px}$  systematically less than the nominal dispersion, and differs by less than $\SI{3}{\micro\eV/px}$ between the two calibrations with a time interval of \SI{4}{\min} \SI{40}{\sec}.
The deviation from the nominal energy solution is dominated by a second-order component common in both exposures, while the third-order component is at the scale of data scattering.
Figure \ref{Fig:2exp}e shows a histogram of the pixel-wise root-mean-square deviation between the energy solutions retrieved from the two exposures, and a normal distribution fit (red curve) indicates a standard deviation of \SI{20.9}{\meV}.
The difference in the calibrated energy dispersions and the standard deviation of the energy solutions are indicative of the `jitter' between two measurements.
The similarity between the two exposures in local irregularities (Fig. \ref{Fig:2exp}a\&e), calibrated dispersion (Fig. \ref{Fig:2exp}b), and global nonlinearity (Fig. \ref{Fig:2exp}c) cross-verifies the accuracy and precision of our frequency-metrology-based calibration method.

%%%%%%%%%%%%%%%%%% Fig. 4 %%%%%%%%%%%%%%%%%%
\begin{figure*}[ht]
\centering
\includegraphics[width=\textwidth]{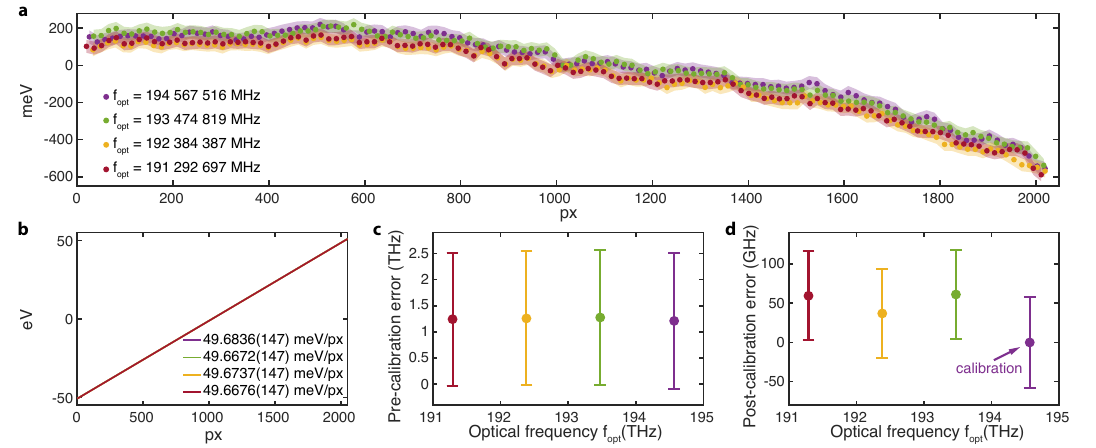}
\caption{
\textbf{Electron spectrometer calibrations with electrons modulated at four different optical frequencies}. 
\textbf{a}, Deviations of the calibration markers at multiples of the photon energy from the nominal energy of the spectrometer. Shaded areas correspond to error bars. 
\textbf{b}, First-order components of the polynomial fits of the energy solutions (pixel-to-energy mappings) for the four calibrations, showing the calibrated average dispersions.
\textbf{c\&d}, Errors in optical frequencies measured from electron spectra before (\textbf{c}) and after (\textbf{d}) calibrating the electron spectrometer with electrons modulated at the optical frequency of \SI{194 567 516}{\MHz}. Error bars in (\textbf{c}) are derived from the maximal deviation of the pre-calibration nominal dispersion from the calibrated dispersion, and error bars in (\textbf{d}) are obtained from the calibration uncertainty of the average dispersion.
}
\label{Fig:4fopt}
\end{figure*}
%%%%%%%%%%%%%%%%%% Fig. 4 %%%%%%%%%%%%%%%%%%
 
We further compared the spectrometer calibrations using electrons modulated at four different absolute optical frequencies: \SI{194 567 516}{\MHz}, \SI{193 474 819}{\MHz}, \SI{192 384 387}{\MHz}, and \SI{191 292 697}{\MHz} (with measurement time intervals of \SI{25}{\min} \SI{25}{\sec}, \SI{18}{\min} \SI{17}{\sec}, and \SI{15}{\min} \SI{19}{\sec}).
Figure \ref{Fig:4fopt}a illustrates the deviation of the calibration markers from the nominal energy for the four calibrations.
The calibrated average dispersions, or the first-order components of the polynomial fits of the energy solutions, are shown in Fig. \ref{Fig:4fopt}b. 
Similarly to the previous case, the calibrated average dispersions reveal a systematic error of $\sim\SI{300}{\micro\eV/px}$ in the nominal dispersion, and differ from each other by less than $\sim\SI{20}{\micro\eV/px}$.

We exhibit an example of electron-based frequency metrology by measuring the optical frequency from the IELS-broadened electron spectrum.
With the number of electron energy sidebands and their energy span, one could back-calculate the photon energy and thus the optical frequency.
Figure \ref{Fig:4fopt}c shows the error in such measurement for the four optical frequencies, using the nominal or pre-calibrated pixel-to-energy mapping of the electron spectrometer.
The errors are in the range of \SI{1.20}{\THz} to \SI{1.28}{\THz}, corresponding to a fractional error of about 6-7 parts in $10^{3}$.
Alternatively, we calibrate the spectrometer using electrons modulated at the optical frequency of \SI{194 567 516}{\MHz}, and the resultant measurement errors in optical frequency become less than \SI{61.1}{\GHz}, corresponding to a fractional error less than 3.2 parts in $10^{4}$.
We emphasize that this value does not represent the precision limit of such measurement, as it does not exclude the effects of instrument performance, instability, and lab environment. 
We also point out that the frequency-resolving capability in our demonstration is directly provided by free electrons, in stark contrast to previous reports where the frequency-resolving capability of high-precision electron spectroscopy comes from tuning the laser frequency~\cite{henkeIntegratedPhotonicsEnables2021c, auadMeVElectronSpectromicroscopy2023}. 

%%%%%%%%%%%%%%%%%%%%%%%%%%%%%%%%%%%%%%%%%%%%%%%%%%%%%%%%%%%%%%%%%%%%%%
%%%%%%%%%%%%%%%%%%%%%%%%%%%%%%%%%%%%%%%%%%%%%%%%%%%%%%%%%%%%%%%%%%%%%%
%%%%%%%%%%%%%%%%%%%%%%%%%%%%%%%%%%%%%%%%%%%%%%%%%%%%%%%%%%%%%%%%%%%%%%
 
\section{Discussion and outlook}
To summarize, we unify frequency metrology across microwave, optical, and free-electron domains.
The cascaded frequency link is enabled by an optical frequency comb that unites microwave and optical frequencies and a photonic chip-based electron phase modulator that bridges optical waves and free electrons.
Our work connects frequency domains differed by a factor of $\sim10^{13}$ and carried by different physical objects - electromagnetic waves and free-electron matter waves.
The precision of such frequency metrology could be further enhanced by state-of-the-art electron monochromators and spectrometers. 
The photonic microresonator-based electron phase modulator provides an efficient electron-photon coupling so that the energy sidebands in the electron spectrum cover the full range of the spectrometer and a single acquisition is sufficient for calibrating the whole spectrometer. 
The efficient coupling is also achieved with both the laser and the electron beam in CW mode, making the calibration technique available in any TEM. 
We anticipate our results would significantly improve the spectral resolution of electron spectroscopy, which already excels at the atomic-scale spatial resolution.
This improvement could further the spectroscopic capabilities of free electrons in vibrational spectroscopy~\cite{krivanekVibrationalSpectroscopyElectron2014}, chemical shift analysis~\cite{tanOxidationStateChemical2012}, quasiparticle search~\cite{husainPinesDemonObserved2023a}, and quantum optics~\cite{feistCavitymediatedElectronphotonPairs2022c}.
Our work sets the milestone for establishing a precise standard or definition of electron energy change in the context of EELS.
The IELS electron spectrum from the photonic chip-based device could calibrate the electron energy-loss spectrum of a specimen when two spectra are taken in succession.   
In-line calibration could also be implemented with a customized sample mount holding both the photonic chip and the specimen. 
Beyond calibration, the investigation lays the foundation for creating an `electron frequency comb' when using photon energy sufficiently larger than the initial electron energy spread, thus enabling novel ultrahigh-precision frequency metrology with free electrons.

\bigskip

%%%%%%%%%%%%%%%%%%%%%%%%%%%%%%%%%%%%%%%%%%%%%%%%%%%%%%%%%%%%%%%%%%%%%%
%%%%%%%%%%%%%%%%%%%%%%%%%%%%%%%%%%%%%%%%%%%%%%%%%%%%%%%%%%%%%%%%%%%%%%
%%%%%%%%%%%%%%%%%%%%%%%%%%%%%%%%%%%%%%%%%%%%%%%%%%%%%%%%%%%%%%%%%%%%%%
\noindent\textbf{Methods}

\smallskip

\begin{footnotesize}

\noindent \textbf{Photonic chip design, fabrication, and packaging}:
The \ce{Si3N4} photonic chip (wafer ID: D66\_01\_F10\_C16) was fabricated by the photonic Damascene process~\cite{pfeifferPhotonicDamasceneProcess2016a, liuHighyieldWaferscaleFabrication2021}, with \ce{SiO2} bottom cladding and no top cladding.
The microresonator has a ring radius of $\SI{20}{\um}$.
The nominal waveguide cross sections are $\SI{2}{\um}\times\SI{650}{\nm}$ and $\SI{800}{\nm}\times\SI{650}{\nm}$ for the microresonator and the bus waveguide, respectively.
The microresonator was characterized by a frequency-comb-assisted broadband optical spectroscopy with tunable diode lasers~\cite{delhayeFrequencyCombAssisted2009b, liuFrequencycombassistedBroadbandPrecision2016}.
The measured total loss rates of the resonances used in the experiment are around $\sim\SI{150}{\MHz}$, corresponding to a quality factor $Q$ of $\sim1.3\times10^6$. 
The input and output fibers were prepared by splicing a short segment of ultrahigh numerical aperture (UNHA-7) fiber to a single mode fiber (SMF-28), and the photonic chip was packaged by attaching the fibers to the bus waveguide inverse taper with an ultraviolet-cured epoxy~\cite{henkeIntegratedPhotonicsEnables2021c}.
The packaged photonic chip was mounted on a custom-built TEM sample holder, and the measured fiber-to-fiber optical transmission is $\sim18.6\%$.

\noindent \textbf{Transmission electron microscopy and electron energy-loss spectroscopy}:
The photonic chip was inserted into a TEM (JEOL JEM-2100PLUS) with a \ce{LaB6} thermionic electron gun. 
The electron beam was emitted by the gun in the continuous-beam mode and accelerated to \SI{120}{\keV}, allowing for phase matching with the microresonator optical modes. 
The beam current was in the range of 5-10 pA. 
The initial energy spread of the electron beam was 0.7-0.8 eV. 
During the experiment, the TEM was in low-magnification ($\times$300 on the control panel) mode with the objective lens off. 
The electron beam was tightly focused close to the surface of the photonic chip with a spot size at the sample of $\sim$40 nm (FWHM) at a distance of $\SI{20}-\SI{30}{\nano\meter}$ from the chip surface, and a convergence angle of $\sim{\SI{100}{\micro\radian}}$ (\SI{50}{\micro\meter} condenser aperture, spot size set with free-lens control of the condenser lenses). 
The electron spectra were measured with a post-column spectrometer (Gatan Imaging Filter GIF Quantum SE) with a 2048$\times$2048 pixels CCD camera (US1000) and $\sim5$ mrad collection semi-angle. Each EELS spectrum was obtained by summing 5 frames of 2-sec exposure each.
The photonic chip was mounted on the tip of a custom-built sample holder with teflon fiber feedthroughs allowing bare optical fibers to enter the TEM column and be coupled to the photonic chip. 
During the experiments, the surface of the photonic chip was kept parallel to the TEM optical axis and to the electron beam optical axis. 
The electron energy-loss spectra were acquired in TEM mode with the electron beam focused tangent to the microresonator waveguide and parallel to its surface.
The electron spectral broadening was maximized by fine-tuning the electron beam position and the photonic chip tilt.

\noindent \textbf{Optical setup}:
The CW laser was emitted by a tunable external cavity diode laser (Toptica) with a wavelength around \SI{1550}{\nm} and amplified by an erbium-doped fiber amplifier (Keopsys). 
The optical power sent to the input fiber was $\SI{50}-\SI{80}{\milli\watt}$. 
The optical polarization was controlled by a fiber polarization controller to excite the transverse magnetic mode family of the microresonator, and the laser frequency was tuned to specific resonant frequencies (shown in Fig. \ref{Fig:4fopt}).
A fraction of the output light was sent to a wavelength meter (HighFinesse) for coarse measurement of the optical frequency.
The wavelength meter was calibrated by an internal neon lamp before each experimental session.
The erbium-doped-fiber-laser-based optical frequency comb (Menlo Systems) was self-referenced and filtered by an optical bandpass filter.
The frequency comb had a repetition rate $f_{\mathrm{rep}}=\SI{251.6575}{\MHz}$ and a carrier-envelope-offset frequency $f_{\mathrm{ceo}}=\SI{20}{\MHz}$.
The optical frequency comb and the output light were mixed at a fast photodiode (New Focus), with their spectral overlap monitored by an optical spectrum analyzer (Yokogawa) and their beatnotes measured by an electronic spectrum analyzer (Rohde \& Schwarz). 
An optical phase-locked loop (OPLL) performed offset locking of the CW laser to one tooth of the optical frequency comb.
Specifically, the lowest-frequency beatnote was brought to $\sim\SI{10}{\MHz}$ by tuning the laser piezo controller.
The beatnote was then filtered, amplified, and compared with the reference, a $\SI{10}{\MHz}$ local oscillator, using a double-balanced 12-bit digital phase detector (Menlo Systems).
The digital phase detector has a large phase detection range and accounts for frequency fluctuation induced phase slips.
The phase detector output, i.e. the error signal, was used as the input signal of a PID servo controller, of which the output signal was fed back to the laser current controller for actuation.
The ambiguity of the beatnote frequency ($\pm$ in Eq. \ref{eq:f_opt}) was removed by observing the shift direction of the beatnote while adjusting the laser frequency when the OPLL was deactivated.
An oscilloscope monitored both the error signal from the phase detector and the optical transmission measured by a photodiode. 

\noindent \textbf{Data processing and analysis}:
The incident electrons, as well as the elastically scattered electrons, have a finite spectral width ($\SI{0.7}-\SI{0.8}{\eV}$) comparable to the photon energy ($\sim\SI{0.8}{\eV}$).
Therefore, the energy sidebands in the IELS electron spectrum are not fully separated and overlap each other.
As a result, the electron spectral peak positions deviate from the regularly spaced energy grid defined by integer multiples of the photon energy.
To mitigate this issue, we first smooth the raw data by convolution with a Gaussian pre-filter, and then use the Richardson-Lucy (RL) deconvolution to retrieve a comb-like spectrum with well-separated peaks~\cite{gloterImprovingEnergyResolution2003}.
The RL deconvolution uses the experimentally measured zero-loss peak (ZLP) spectrum as the deconvolution Kernel, after proper normalization and zero-padding.
For electron spectrometer calibration, the peak positions of the deconvolved, comb-like electron spectrum are used as calibration markers.
Pixel positions (with sub-pixel precision) of the calibration markers are assigned an energy of the corresponding integer multiples of the photon energy, which is determined from absolute optical frequency measurement.
Pixels in between the calibration markers are assigned an energy from linear interpolation.
As the result, the energy solution, i.e. the pixel-to-energy mapping, of the spectrometer is retrieved.
The energy solution is then fitted with a third-order polynomial, with the slope of the linear component representing the calibrated average dispersion.
The iterative RL deconvolution typically has an optimal number of iterations, at which point the deconvolution has to be terminated to prevent noise amplification~\cite{biggsAccelerationIterativeImage1997}.
For each electron spectrum, the deconvolution is tested for a range of parameter combinations with the Gaussian pre-filter width in 0.01-0.16 eV (standard deviation) and the number of iterations in 10-300.
The optimal parameter set is determined by evaluating the calibration results, i.e. the calibrated average dispersion and energy solution, in three validation methods: (1) simulation-validation, where the deconvolution is performed on a simulated IELS electron spectrum with a known spectrometer energy solution as the ground truth; (2) self-validation, where a single experimental electron spectrum is deconvolved with different parameter combinations; (3) cross-validation, where two experimental electron spectra generated with the same absolute optical frequency and recorded in rapid succession (less than \SI{5}{\min} time interval) are deconvolved and compared.
In consequence, the obtained range of optimal parameter combinations is 0.05-0.09 eV for the Gaussian pre-filter width and 60-100 for the number of iterations.
The same range of parameter combinations is applied in the processing of all electron spectra.
We use a range rather than a single parameter combination to ensure the robustness of the calibration method and to estimate the calibration uncertainty.
The final calibration is the average of the calibrations obtained from all parameter combinations in the aforementioned range, and the maximal deviation from the average is taken as the uncertainty, which is \SI{14.68}{\micro\eV/px} in the average dispersion and \SI{35.08}{\meV} (0.7 pixel under nominal dispersion) pixel-wise root-mean-square uncertainty in the energy solution. 
These two uncertainty values are obtained from cross-validation and are the largest values from all three validation methods.
Hence, the two uncertainty values represent a conservative estimate and do not exclude short-term jitter from instrument drift, instability, and lab environment. 

\medskip

\noindent \textbf{Acknowledgments}:
All photonic integrated circuit samples were fabricated in the Center of MicroNanoTechnology (CMi). 
The TEM experiments were conducted at the Lausanne Center for Ultrafast Science (LACUS).
We thank J. Riemensberger and G. Likhachev for help with the optical frequency comb, G. Huang and N. Engelsen for help with the wavelength meter, A. Davydova for assistance with data acquisition, and T. Bl\'esin for taking the optical micrograph.
This material is based on work supported by the Air Force Office of Scientific Research under award FA9550-19-1-0250, the ERC consolidator grant 771346 (ISCQuM), the EU H2020 research and innovation program under grant agreement 964591 (SMART-electron), and EPFL Science Seed Fund. 
Y.Y. and B.W. acknowledge support from the EU H2020 research and innovation program under the Marie Skłodowska-Curie Actions grant agreement 101033593 (SEPhIM) and grant agreement 801459 (FP-RESOMUS) respectively. 

\medskip

\noindent \textbf{Author contribution}:
Conceptualization, Supervision, Project administration, Funding acquisition: T.J.K., T.L., and F.C.
Methodology: Y.Y., P.C., A.S.R., B.W., R.N.W., A.S., and T.L.
Software: Y.Y.
Validation: Y.Y. and P.C.
Formal analysis: Y.Y., P.C., and T.L.
Investigation: Y.Y., P.C., B.W., and T.L.
Resources: A.S.R., R.N.W., and T.L.
Data curation: Y.Y. and P.C.
Visualization: Y.Y. and P.C.
Writing (original draft preparation): Y.Y. and T.J.K.
Writing (review and editing): all authors.

\medskip

\noindent \textbf{Data Availability Statement}: The code and data used to produce the plots within this work will be released on the repository \texttt{Zenodo} upon publication of this preprint.

\end{footnotesize}

\pretolerance=0
\bigskip
\bibliographystyle{apsrev4-2}
\bibliography{bibliography}
\end{document}